\documentclass[preprint,showpacs,preprintnumbers,amsmath,amssymb]{revtex4}

\usepackage{graphicx}
\usepackage{dcolumn}
\usepackage{bm}

\newcommand{\ber} {\begin {eqnarray}}
\newcommand{\enr} {\end {eqnarray}}

\newcommand {\SE} {Schr\"{o}dinger equation}

\newcommand {\er}[1] {equation (\ref{#1}) }

\begin{document}

\title{ Phase Evolution in a Multi-Component System}

\author{Robert Englman}
\email{englman@vms.huji.ac.il}
 \affiliation{Department of Physics and Applied Mathematics,\\ Soreq NRC,Yavne
81800,Israel}
 \altaffiliation[and ]{College of Judea and Samaria, Ariel 44284, Israel}
\author{Asher Yahalom}
 \email{asya@ycariel.yosh.ac.il}
\affiliation{College of Judea and Samaria, Ariel 44284, Israel}

\date{\today}

\begin {abstract}
We derive a general expression  for the expectation value of the
phase acquired by a time dependent wave function in a multi
component system, as excursions  are made in its coordinate space.
We then obtain the mean phase for the (linear dynamic $E \otimes
\epsilon$) Jahn-Teller situation
  in an electronically degenerate system. We interpret the phase-change
  as an observable measure of the {\it effective} nodal structure of the
   wave function.

\end {abstract}

\pacs{03.65.Vf, 31.50.Gh, 31.30.Gs, 71.70.Ej} \maketitle

\maketitle

 In a recent publication a geometric (or
Berry) phase was calculated for the wave function of a multi-component
closed system \cite {FuentesGCBV}. This differs from the usually
considered situations, in which  the Berry phases emerge from the
wave-function of the system during the cyclic evolution of some external parameter.
 It is of interest to point out that in a famous prototype of a multi-component,
  closed system (an electronically doubly degenerate molecule) dynamic
solutions for the $E \otimes \epsilon$ linear Jahn-Teller
 effect (DJTE) were fully obtained as long ago as
1957 \cite{MoffitT, LonguetOPS}. The two parts of
system were the electronic and ionic constituents of the molecule.
 Though this has  received, as just noted, a
complete treatment early on, the dynamic problem has not left the
scientific agenda ever since. Descriptions of some of the early
refinements are found in two books \cite {Englman, BersukerP};  the most
recent publication known to us and involving a variational treatment of
the problem  is in \cite{DunnE}. The physical consequences of the Berry
phase on the DJTE were clearly brought out by the late Frank Ham
\cite{Ham} and more recently in \cite{KoizumiB}, both of which papers
showed (albeit under different physical conditions) that the value of the
Berry phase may be critical in determining the order of energy levels in
the closed molecular system. This phase is thus clearly observable by
experiment \cite{Ham}. Its physical interpretation, essentially along the
lines of \cite {Ham,KoizumiB}, will be given later in this work.

In \cite {FuentesGCBV} an operator was proposed for the phase change
 (called "quantized phase") in a closed system. Here we shall derive an
expression for this phase from first principles and use it to calculate
the phase change in the vibronic doublet ground states of an $E \otimes
\epsilon$ linearly coupled Jahn-Teller system.  We shall use a
"guessed solution", for the ground state, which is transparent,
intuitively simple and algebraically easily manageable \cite{Englman62,
EnglmanH, Englman}. Though not variationally obtained, the "guessed
solution" was found to have eigen-energies that are considerably closer
to the exact, computed energies of \cite{LonguetOPS} than any other
approximate solution with which it was compared. This comparison is seen
in Fig. 2 of \cite{ZhengB}. Later treatments did not test their methods
by comparison with the "guessed solution".

 Our point of departure is the time-dependent \SE
\begin{equation}
i\frac{\partial}{\partial t} \psi({\bf r},t) = H({\bf r},t)\psi({\bf r},t)
\label {tdse1d}
\end{equation}
 for a wave function $\psi({\bf r},t)$ that
depends on the internal coordinates ${\bf r}$ of the system, as well as
on time $t$. ($\hbar=1$) The system is coupled to the environment; hence
the dependence on $t$ in the Hamiltonian. In a  closed system, the
Hamiltonian is time-independent. $\psi$ is still time-dependent, as,
e.g., in a wave-packet.

We write the state (assumed to be regular in the coordinate space and
vanishing at its boundaries) as
\begin{equation}
\psi=A e^{iS}
\label{psidirac}
\end{equation}
with $A$ and $S$ being real functions of ${\bf
r},t$. ($A>0$.) We shall utilize  the equation of continuity and the
Hamilton-Jacobi equation \cite{Messiah}:
\begin{equation}
\frac{\partial A^2}{\partial t} =-\frac{1}{m}\nabla( A^2\nabla S)
\label {eoc}
\end{equation}
\begin{equation}
\frac{1}{2m}(\nabla S)^2= -\frac{\partial S}{\partial t} - V({\bf r},t)
+\frac{1}{2m}A^{-1}(\nabla)^2 A
\label{HJe}
\end{equation}
Here $m$ is a  mass parameter common to all
degrees of freedom, with all coordinates scaled to this mass. $V$ is the
potential.

 Let us now consider the change in the wave
function, between the initial state of the system  at $t=0$ and
 a final time  $t_f$. Real and imaginary parts of the change in the logarithm are \begin{equation}
 [ln\psi]^{t_f}_0 = [ln|\psi|]^{t_f}_0 +i[arg(\psi)]^{t_f}_0= [ln A]^{t_f}_0
  +i[S]^{t_f}_0
 \label{tchange1}
 \end{equation}
These are functions of the coordinates. To form quantum mechanical
expectation values (denoted by angular brackets about the relevant
quantities) we multiply by $A^2$ and integrate over all coordinates. Thus
the mean  of the changes can be written as \begin{equation} <[ln A]^{t_f}_0 >
  +i<[S]^{t_f}_0> = \int_0^{t_f} dt \frac{\partial}{\partial t} \int d{\bf r}
A^2(lnA +iS)\label {tchange2}\end{equation} [ It is natural to conjecture
that to form statistical expectation values (appropriate to mixed
states \cite{SjoqvistPEAEOV}), the factor $A^2$ is to be
multiplied by the relative statistical weight of the state and the
contribution due to all states be summed over. We do not pursue
this topic here.] Separating real and imaginary parts we get for
the real part the quantity $ -\frac{1}{2} [{{\cal S}_e}]^{t_f}_0 $
where  \begin{equation} {{\cal S}_e}
   =-\int d{\bf r} |\psi |^2 ln |\psi |^2 \label {entropy}\end{equation}  ${\cal S}_e$ (which
is different from the phase $S$) is reminiscent of a von Neumann entropy,
in which the  density operator is projected onto the initial state.

 We turn now to the rate of change of the expectation value of the phase
 $\frac{d<S>}{dt}$. We change the order of integrations in \er {tchange2} and obtain
 after some manipulations :
   \ber
    \frac{d<S>}{dt} &=& \int d{\bf r}( A^2\frac{\partial S}{\partial t}~+~
   S\frac{\partial A^2}{\partial t}) = \int d{\bf r}[ A^2\frac
   {\partial S}{\partial t} ~-~\frac {1}{m} S\nabla (A^2 \nabla
   S)]\nonumber\\ & = & \int d{\bf r}[ A^2\frac
   {\partial S}{\partial t}+ \frac {1}{m} A^2 (\nabla S)^2]\label
   {imtchange}
   \enr having used \er {eoc} and integrated by parts (with
   vanishing integrands at space-extremities). We now substitute for $(\nabla
   S)^2$ from \er{HJe} and obtain a  change in the sign of the first term
    in the above
    expression, as well as the expectation values of (twice) the
    potential and a term related to the kinetic energies, which can be
 reworked by a further integration by parts so as to put it into a form of
 definite sign, giving
 \begin{equation}
 \frac{d<S>}{dt} = - \int d{\bf r}( A^2\frac{\partial S}{\partial t})~-~
    2<V({\bf r},t)> ~-~\frac{1}{m}\int d{\bf r}(\nabla A)^2
    \label{imtchange2}\end{equation} We next recall that $S=Im (ln \psi)$ and
 reinstate the time integration to get the change of phase as
 \ber
 [<S>]^{t_f}_0  &=& -\int^{t_f}_0 dt \int d{\bf r}[Im (\psi^+({\bf r},t)
 \frac{\partial \psi({\bf r},t)}{\partial t}) + 2|(\psi({\bf r},t)|^2 V({\bf r},t)
 \nonumber\\
  &+& \frac{1}{m} (\nabla |\psi({\bf r},t)|)^2]
  \label{imtchange3}
 \enr
     where the cross means Hermitian conjugate. On the other hand,
     multiplying \er{tdse1d} by $\psi^+({\bf r},t)$, integrating over the
     coordinates and again integrating by parts, we obtain
  \begin{equation}
  Im\int d{\bf r}(\psi^+({\bf r},t)
    \frac{\partial
   \psi({\bf r},t)}{\partial t})=-\int d{\bf r}
    \frac{|\nabla \psi({\bf r},t)|^2}{2 m} - \int d{\bf r} |(\psi({\bf r},t)|^2
    V({\bf r},t) \label{roi}\end{equation} which we use to eliminate the
    expectation value of the potential from \er{imtchange3}. We then get:
\ber
[<S>]^{t_f}_0 & = &Im\int^{t_f}_0 dt \int d{\bf r}(\psi^+({\bf r},t)
    \frac{\partial \psi({\bf r},t)}{\partial t})\nonumber\\ &+&
    \frac{1}{m}  \int^{t_f}_0 dt
     \int d{\bf r}
    [|\nabla \psi({\bf
    r},t)|^2-(\nabla |\psi({\bf r},t)|)^2]\label{imtchange4}\enr

     This is our central result for the mean phase change. The first term is of the form familiar in
     e.g., expressions of the open path geometric phase \cite {PatiJ}.
     In the second term, to be denoted for brevity $\delta K $, one has
      the difference between two space-derivative
     terms, one involving the total (complex) time-dependent wave-function
     and the other its modulus.

We now calculate the phase change in our molecular model for a
multi-component closed system. The main simplification in the model is
the restriction to a two-dimensional electronic subspace (it being assumed
 that other electronic states of
the molecule are too far away to have any effect) and small
displacements of the nuclear coordinates from some standard configuration
(so that only linear terms in the nuclear displacement coordinates appear
in the Hamiltonian below). The solution to the mathematical problem (the DJTE)
 embodies the correlated nuclear-electronic
trajectory near a conical intersection of the (diabatic) potential
surfaces. Under these circumstances the usual Born-Oppenheimer approximation
breaks down and the description  of the combined dynamics is non-trivial.

 The total Hamiltonian consists of $H_{mol}$ for the internal
  degrees of freedom of the molecule
and an interaction term with the environment $H_{env}$.
 \begin{equation} H = H_{mol}+ H_{env} \label {H1}\end{equation}
 The first term is a function of the electronic and nuclear coordinates, while the
second term may contain also an externally imposed time dependence. Our
 restriction to a two-dimensional electronic subspace $|1>, |2 >$
 removes from the formalism the presence of electronic coordinates and
leaves only the nuclear coordinates. Two of these, designated $q_a, q_b$,
 are of interest. $H_{mol}$ when expressed in terms of the bosonic creation $(a^+ ,b^+)$ and
 annihilation $(a ,b)$ operators of the nuclear motion, takes the  form
 \begin{equation} H_{mol}= \frac{\omega}{2}\{a^+a +~b^+b  -\frac
{k}{\sqrt{2}}[(a^+ + a)\sigma_z - (b^+ + b)\sigma_x]\}
\label{H2}\end{equation} Here $\omega$ is the frequency of oscillation of the
nuclear motion, and $k$ is the electron-nuclear coupling strength expressed
 in dimensionless units.
The $2$ x $2$ matrices $\sigma_x$ and $\sigma_z$ are the familiar
Pauli operators acting on the electronic $|1>,|1>$ subspace. Equivalent
 representations of the Hamiltonian $H_{mol}$
are given in works on the Jahn-Teller effect (\cite
{Englman62,Englman,BersukerP}); i.e., in terms of the nuclear coordinates
$q_a, q_b$ or of the associated cylindrical coordinates $(q,\phi)$, where
 $q_a= qcos\phi, ~q_b=qsin\phi$, as, e.g., in  Eq.
3.5 of \cite {Englman}.

   The algebraic expression for the
   ground state doublet proposed in \cite {Englman62,EnglmanH}, and which
  solves the  time-independent \SE~for the Hamiltonian $H_{mol}$ to a good
   approximation,  has the
   following (unnormalized) form:\begin{equation} \hat\psi(q_a,q_b)= \exp-\frac{1}{2}
   \big[(q_a- k\sigma_z)^2 +(q_b+ k\sigma_x)^2]\label {psi1}\end{equation}
(Intuitively, this form is suggested by analogy with the ground state
solutions of displaced harmonic oscillators, but its justification is in
its close agreement with exactly computed eigenvalues of $H_{mol}$
\cite{LonguetOPS}). To obtain the ground state doublet we operate with
$\hat\psi(q_a,q_b)$ on any two linearly independent combinations of the
basic vectors $|1>,|2>$. In a column vector representation
these are just $  {\tiny \left(
\begin{array}{cc}
1 \\ 0 \end{array} \right),~\left(
\begin{array}{cc}
0 \\ 1 \end{array} \right)\label {vectors} }$. The exponential,
which includes non-commuting matrices, can be manipulated by use
of
 the  commutation relations between the Pauli-matrices to give, in terms
of the cylindrical coordinates defined above,  the expression
\begin{equation}
\hat\psi (q,\phi)=\exp[-k^2-q^2/2 ][\cosh(kq){\bf I}
-\sinh(kq)(\sigma_z \cos\phi -\sigma_x \sin\phi)]\label {psi2}
\end{equation}
${\bf
I}$ is the unit $2$ x $2$ matrix. One notes that this is a single valued
function of $\phi$ (there are no $\cos\phi/2$ terms), as indeed is
required of the wave function of a closed system \cite {Ham}.

By operating with $\hat\psi$ on $ {\frac{1}{\sqrt 2}}\left(
\begin{array}{cc}
1 \\{\mp} i \end{array} \right) $ one gets the two degenerate ground-
state functions
 $\Psi(q,\phi)(=\Psi_-)$ and $\Psi^*(q,\phi)(=\Psi_+)$. The eigenvalues and other
 related properties of these states have been calculated in \cite{EnglmanH,
  Englman}. Here we
 compute the  phase-change for each function, as the angular
 coordinate  changes by a full period between $\phi=0$ and $\phi=2\pi$.

 One procedure to induce such change in an internal coordinate (and
  physically, perhaps, the only consistent one) is to consider it being
guided by an external force along
   a circle. (The concept of a guiding potential was used in,
  e.g, \cite{AharonovB}, but here we guide the angular coordinate, rather than the
  radial one.) To achieve this, one needs an external, time dependent
  agent acting on the otherwise
 closed system and this is the role played by $H_{env}$ in the Hamiltonian
  shown in \er{H1}.  We suppose that there is an environment Hamiltonian which
  induces  a $delta$-function like behavior in the wave function
(forcing $\phi$ to equal $\Omega t$) and that this time-dependent
 $H_{env}$ dominates
 the kinetic energy of the angular variable. In this, delta-function limit the  variable
 $ \phi$ turns into a (classical) parameter and is no
 longer a "degree of freedom".

 We thus get from \er{imtchange4} for the expectation value of the phase
  change in the $\Psi^-$ state the expression
  \ber
  <[S]^{t_f}_0>
 &=& [Im \int^{t_f}_0 dt[ \int^{\infty}_0 dq q \Psi^+_-(q,\Omega t)
 \frac{\partial}{\partial t}\Psi_{-}(q,\Omega t)
\nonumber \\
 &+& \delta K]/\int^{\infty}_0 dq q |\Psi_{-}(q,\Omega t)|^2]
 \label{gammaminus}
 \enr
 where $\delta K$ is the second term on the
 right hand side of  \er{imtchange4}
 involving the space differentials. In all integrals the
  {\it integration is over the radial coordinate $q$ only,
   since $\phi=\Omega t$ is treated as a parameter}.

\begin{figure}
\vspace{7cm} \includegraphics{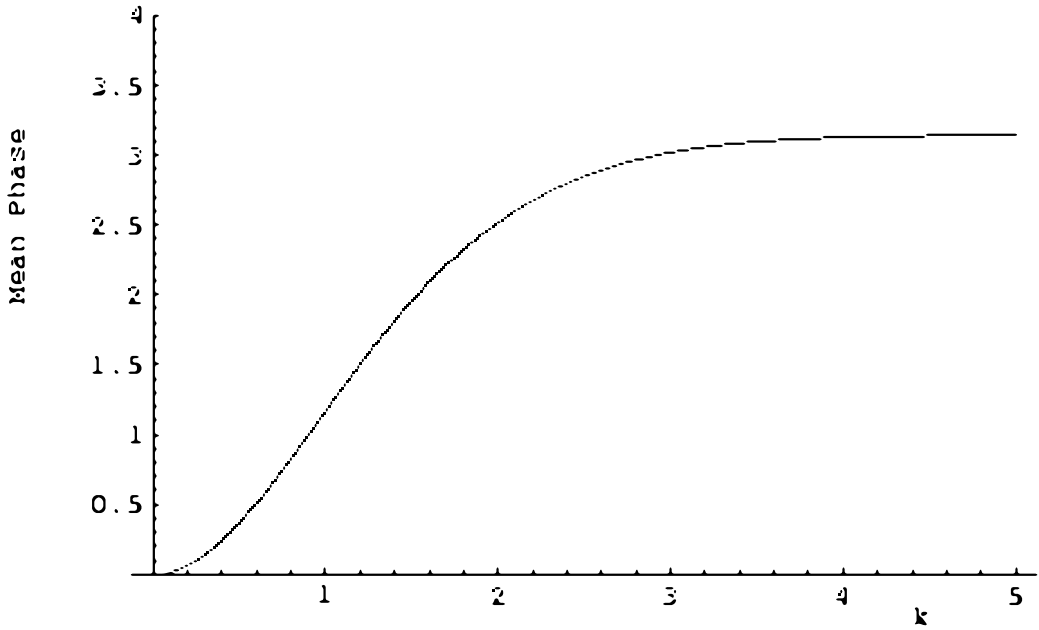} \caption{Expectation value of the phase
change after a full cycle in the coordinate space {\it versus} the
coupling strength. We plot  $ [<S>]^{2\pi /\Omega}_0$ (
\er{gammaminus}, with $\delta K=0$ ) against k (introduced in \er
{H1}). For large $k$, the phase approaches $\pi$.}
\label{averagphase}
\end{figure}

It can be shown that $\delta K $ is identically zero for the both $\Psi_{-}$
 and $\Psi_{+}$. (Remember
that the gradient operator in $\delta K$ involves now only the radial
degree of freedom $q$.) The evaluation of the first integral in
\er{gammaminus} leads to the plot shown in the following figure for the
mean phase after a full cyclic revolution $<[S]^{2\pi /\Omega}_0>$ as
function of the coupling strength $k$. (Fig. \ref{averagphase})

 As seen in the figure, the acquired mean phase for $\Psi_{-}$
  increases monotonically with
 the coupling and levels off for strong coupling $(k>>1)$ to
 $\pi$ (the value of the Berry-phase). The corresponding
 phase for the partner state $\Psi_{+}$
 is the negative of this value, and any linear combination of the  ground state
 doublet will result in intermediate values between the two extremes $\pm
 \pi$.  The phase
   depends only on the strength of the coupling $k$. It is independent of the
   adiabatic parameter $\frac{k\omega}{\Omega}$, since the
   integrand contains only the instantaneous value of the initial component $\Psi_{-}$ ,
    and no admixture from its partner $\Psi_{+}$ .
(Applying our \er{gammaminus} to the eigenstates $\Phi_n^{\pm}$ of \cite
{FuentesGCBV} expressed in a coordinate representation reproduces exactly the
results obtained in that  paper. However, evaluation of the expectation value of
the phase-shift operator
   proposed in \cite {FuentesGCBV} for the states $\Psi^{\pm}(q,\phi)$
   in this work, where the rotating-wave approximation is not made,
   yields values that diverge quadratically for large k.)

  We conclude with an interpretation of the "closed-system" phase. In  this
 we follow  \cite {Ham, KoizumiB}.  For low
 values of the coupling  constant, the wave function is smeared over the
 origin $q=0$ and cannot be said to circle {\it around} this point, which is a
 point of degeneracy of the two states. Then there is hardly any  acquired
  phase. For large
  values of the coupling, the wave function is located near $q=k$, meaning
  that it keeps away from origin, so that circling ${\it around}$ it  can
achieve the  full measure of the geometric phase.

  On the other hand, it has been known for some time that (in the
  adiabatic limit) the phase change comes about abruptly, precisely at the
  moment of circling when  a component amplitude vanishes. (This occurs when
$cos \frac{
  \Omega t}{2}=0$, or $\Omega t =\pi$. The abrupt change is clearly seen
 in the figures
   of \cite{EnglmanY1999, EnglmanYB2000} and has  recently formed the subject
 of a Letter  \cite {Sjoqvist}.)
  By the interpretation just given, the phase change is a
 measure of the extent that
  a circling  in the coordinate space scans  the zeros of the wave function  in
the region encircled. Since zeros (nodes) in the wave function are known to
  affect (in general, raise) the energies of the states, it is natural to find
    that the phase acquired during a revolution determines
    the ordering  of the energy levels. Such connections between phase change
    and energy levels  have been  noted first in \cite {Ham}
    and more recently in \cite { KoizumiB}.

\begin{acknowledgments}
 We thank  Prof. Roi Baer for helpful discussions.
\end{acknowledgments}

\end {document}